\pdfoutput=1
\documentclass[12pt]{scrartcl}
\usepackage{hyphenat}

\usepackage[T1]{fontenc}         
\usepackage[utf8]{inputenc}
\usepackage[british]{babel}
\areaset{158mm}{238mm}
\usepackage{setspace}
\usepackage[all]{xypic}
\usepackage{scrlayer-scrpage}
\usepackage{tikz}
\usetikzlibrary{cd}

\usepackage{amsmath}
\usepackage{mathtools}
\usepackage{mathrsfs}
\usepackage{physics}

\usepackage{amssymb}
\usepackage[mathscr]{eucal}
\usepackage{cite}
\usepackage{bm}
\usepackage{slashed}

\usepackage{tikz,pgf}
\usetikzlibrary{shapes}
\usetikzlibrary{calc}
\usetikzlibrary{decorations.pathmorphing}
\usetikzlibrary{decorations.pathreplacing,shapes.misc}
\usetikzlibrary{positioning}
\usetikzlibrary{decorations.markings}
\tikzset{->-/.style={decoration={
  markings,
  mark=at position .5 with {\arrow{>}}},postaction={decorate}}}
\tikzset{-<-/.style={decoration={
  markings,
  mark=at position .5 with {\arrow{<}}},postaction={decorate}}}

\newcommand{\be}{\begin{equation}}
\newcommand{\ee}{\end{equation}}

\usepackage[unicode=true,linktoc=all]{hyperref}

\numberwithin{equation}{section}

\title{\vspace{-1cm} On the 6d Origin of Non-invertible Symmetries in 4d}

\author{Vladimir Bashmakov$^{\dagger}$, Michele Del Zotto$^{\dagger\sharp}$, \\ and Azeem Hasan$^{\sharp}$
\\[1cm]
	\small\slshape$^\sharp$ Mathematics Institute, Uppsala University,  \\[-0.2cm] 
	\small\slshape Box 480, SE-75106 Uppsala, Sweden\\
	\small\slshape$^\dagger$ Department of Physics and Astronomy, Uppsala University,  \\[-0.2cm] 
	\small\slshape Box 516, SE-75120 Uppsala, Sweden\\
	}
\date{}

\begin{document}

\maketitle

\paragraph{\hspace{.9cm}\large{Abstract}}
\vspace{-1cm}
\begin{abstract}

\noindent It is well-known that six-dimensional superconformal field theories can be exploited to unravel interesting features of lower-dimensional theories obtained via compactifications. In this short note we discuss a new application of 6d (2,0) theories in constructing 4d theories with Kramers-Wannier-like non-invertible symmetries. Our methods allow to recover previously known results, as well as to exhibit infinitely many new examples of four dimensional theories with ``M-ality" defects (arising from operations of order $M$ generalizing dualities). In particular, we obtain examples of order $M=p^k$, where $p>1$ is a prime number and $k$ is a positive integer.

\end{abstract}

\vfill{}
--------------------------

\today 
\thispagestyle{empty}

\newpage

\tableofcontents

\section{Introduction}

Our understanding of generalizations of global symmetries, which capture the quantum numbers of extended operators in quantum fields
\cite{Gaiotto:2010be,Kapustin:2013qsa,Kapustin:2013uxa,Aharony:2013hda,Gaiotto:2014kfa}, is undergoing a rapid evolution
\cite{DelZotto:2015isa,Sharpe:2015mja,Tachikawa:2017gyf,Cordova:2018cvg,Benini:2018reh,Hsin:2018vcg,Thorngren:2019iar,GarciaEtxebarria:2019caf,Eckhard:2019jgg,Bergman:2020ifi,Morrison:2020ool,Albertini:2020mdx,Hsin:2020nts,Bah:2020uev,DelZotto:2020esg,Hason:2020yqf,Bhardwaj:2020phs,Apruzzi:2020zot,Cordova:2020tij,Thorngren:2020aph,DelZotto:2020sop,BenettiGenolini:2020doj,Yu:2020twi,Bhardwaj:2020ymp,DeWolfe:2020uzb,Gukov:2020btk,Iqbal:2020lrt,Hidaka:2020izy,Brennan:2020ehu,Komargodski:2020mxz,Closset:2020afy,Thorngren:2020yht,Closset:2020scj,Bhardwaj:2021pfz,Nguyen:2021naa,Heidenreich:2021xpr,Apruzzi:2021phx,Apruzzi:2021vcu,Hosseini:2021ged,Cvetic:2021sxm,Buican:2021xhs,Bhardwaj:2021zrt,Iqbal:2021rkn,Braun:2021sex,Cvetic:2021maf,Closset:2021lhd,Thorngren:2021yso,Sharpe:2021srf,Bhardwaj:2021wif,Hidaka:2021mml,Lee:2021obi,Lee:2021crt,Hidaka:2021kkf,Koide:2021zxj,Apruzzi:2021mlh,Kaidi:2021xfk,Choi:2021kmx,Bah:2021brs,Gukov:2021swm,Closset:2021lwy,Yu:2021zmu,Apruzzi:2021nmk,Beratto:2021xmn,Bhardwaj:2021mzl,Cvetic:2022uuu,DelZotto:2022ras,DelZotto:2022joo,DelZotto:2022fnw,Bhardwaj:2022yxj,Hayashi:2022fkw,Kaidi:2022uux,Roumpedakis:2022aik,Choi:2022jqy,Choi:2022zal,Arias-Tamargo:2022nlf,Cordova:2022ieu,Bhardwaj:2022dyt,Benedetti:2022zbb,Bhardwaj:2022scy,Antinucci:2022eat,Cvetic:2022imb,Wan:2018bns,Wan:2019soo,Wang:2021vki,Carta:2022spy}.
Symmetries of quantum fields are understood in terms of subsectors of extended quasi-topological defect operators whose fusion rules generalize the notion of
groups, but can be characterized, exploiting higher categories, thus producing so-called \textit{global categorical symmetries}. In particular, such generalized
symmetry defects are not necessarily invertible \cite{Thorngren:2019iar,Komargodski:2020mxz}.

\medskip

While such non-invertible symmetries in $1+1$-dimensional QFTs are well known (see e.g. the discussion in \cite{Dijkgraaf:1989hb}), it is a more recent result that 3+1 dimensional QFTs admit non-invertible symmetry defects as well \cite{Kaidi:2021xfk,Choi:2021kmx} (see also \cite{Bhardwaj:2022yxj,Choi:2022zal,Cordova:2022ieu}). One can formulate two (pretty much related) strategies to produce such examples. The first builds on a generalization of the Kramers-Wannier duality defect, and involves finding models with a self-dual point in their moduli spaces, so that the duality defects become symmetries for these systems, but with fusion rules happening to be non-invertible \cite{Choi:2021kmx,Choi:2022zal,Cordova:2022ieu}. For the second, given a theory $\mathcal T$ with a $\mathbb Z_M^{(0)} \times \mathbb Z_N^{(1)}$ symmetry with a mixed anomaly, one constructs a new theory by gauging the $\mathbb Z_N^{(1)}$ symmetry: the resulting theory has non-invertible codimension-one symmetry defects, descending from the original symmetry $\mathbb{Z}_M^{(0)}$. The corresponding fusion rules can be explicitly computed as a result of this construction \cite{Kaidi:2021xfk}. This short note is the first in a series, whose purpose is to describe simple applications of 6d (2,0) superconformal field theories (SCFTs) in producing infinitely many examples of 3+1 dimensional theories with non-invertible global categorical symmetries of these kinds. This is yet another of the many applications of higher-dimensional SCFTs in unravelling interesting features of lower-dimensional dynamics.

\medskip

More precisely, our technique can be explained as follows. 6d (2,0) SCFTs compactified on a Riemann surface $\Sigma_g$ (possibly with punctures, suitably chosen and decorated) give rise to 4d $\mathcal N=2$ SCFTs, whose conformal manifolds are identified with the moduli spaces of complex structures of $\Sigma_g$; these theories are known in the literature as theories of class $\mathcal S$  \cite{Gaiotto:2009hg,Gaiotto:2009we}. In particular, the S-duality group of these models is identified with the mapping class group of $\Sigma_g$ \cite{Gaiotto:2009we}. Since the action of the mapping class group is not free, there are points which admit a non-trivial stabilizer: the corresponding class $\mathcal S$ theories have an enhanced 0-form symmetry provided their global structure is compatible with such an action --- the stabilizer subgroup of the S-duality group becomes an ordinary 0-form symmetry for the theory corresponding to that point in moduli space. Since S-duality transformations typically rotate the global structure of the theory of interest, the resulting 0-form symmetry is expected to have a mixed anomaly with the corresponding 1-form symmetries. One can explicitly detect these mixed anomalies, exploiting the relative nature of 6d (2,0) SCFTs and its interplay with the global structure of class $\mathcal S$ theories \cite{Tachikawa:2013hya}. This gives a geometric origin for a large class of non-invertible symmetries in 4d theories with a mixed anomaly origin.

Whenever the stabilizer subgroup of the S-duality does not respect the global structure, we can compensate its action via gauging of a subgroup of the
one-form symmetry, thus giving rise to intrinsic Kramers-Wannier duality defects, which do not have a mixed anomaly origin. In this paper we focus on cases that exhibit a mixed anomaly. In a follow up of this work we will study the more general case of intrinsic duality defects, building on uplifting the formalism introduced in \cite{Kaidi:2022uux} to our setup.

\medskip

The power of the 6d approach is clear from the outset: our results can be obtained in few lines, starting from the simplest possible examples of class $\mathcal S$ theories, arising from the compactification of 6d (2,0) SCFTs of type $A_{n-1}$ on Riemann surfaces $\Sigma_g$ of genus $g$ without punctures. As an example of the power of this method, we showcase an infinity of theories with non-invertible symmetries of orders $M=p^k$, where $p>1$ is a prime number and $k$ is a positive integer.

\medskip

The structure of this note is as follows. In section \ref{sec:MALITY} we review the argument of \cite{Kaidi:2021xfk} for obtaining Kramers-Wannier-like non-invertible symmetries in four-dimensional theories. In section \ref{sec:global6d}, to establish some notations and conventions, we quickly review the Tachikawa's method for
reading off the global structures of class $\mathcal S$ theories from 6d  \cite{Tachikawa:2013hya}. In section \ref{sec:strata} we discuss our general strategy in more detail, formulating a sufficient criterion for the existence of a mixed anomaly. In section \ref{sec:ex} we discuss several applications of our method. In particular, in \S.\ref{sec:an-example:-mixed} we rederive the original $\mathcal{N}=4$ example in \cite{Kaidi:2021xfk} from the 6d perspective, in \S.\ref{sec:Nequal4} we give further 4d $\mathcal N=4$ examples, and in \S.\ref{sec:infinitelymany} we construct infinitely many examples of models with non-invertible symmetry defects of orders $M=p^k$, where $p>1$ is a prime number and $k$ is a positive integer. In section \ref{sec:conclusions} we present our conclusions and outlook. An alternative strategy to produce other kinds of non-invertible symmetry defects is briefly sketched in the appendix, building on the mechanism illustrated in \cite{Bhardwaj:2022yxj}.

\section{``M-ality'' from $\mathbb Z_M^{(0)} \times \mathbb Z_N^{(1)}$ mixed anomalies}\label{sec:MALITY}

In this section we review the construction of \cite{Kaidi:2021xfk} for generating non-invertible $M$-ality defects, starting from a 4d theory $\mathcal T$ with a $\mathbb Z^{(0)}_M \times \mathbb Z^{(1)}_N$ mixed anomaly.\footnote{\ The main emphasis in  \cite{Kaidi:2021xfk} is on duality defects: the $M$-ality case is somewhat implicit in the appendix of that paper. We are thankful to Justin Kaidi for sharing his insight on this more general argument with us.} 

\medskip

Consider coupling $\mathcal T$ to background gauge fields for the  $\mathbb Z_M^{(0)} \times \mathbb Z_N^{(1)}$ symmetry, which we denote by $A^{(1)}$ and $B^{(2)}$. Assume $\mathcal T$ has a mixed anomaly
\be
Z_{\mathcal T}[A^{(1)}+d \lambda^{(0)},B^{(2)}] = e^{i{2 \pi \over 2 N} p \int_{X}  \lambda^{(0)} \mathcal{P}(B^{(2)}) }Z_{\mathcal T}[A^{(1)},B^{(2)}],
\ee
where $p$ is an integer. Let us denote the codimension one topological defects associated to the $\mathbb Z_M^{(0)}$ symmetry  $D_3(M_3,B^{(2)})$, where we are emphasizing their explicit dependence on $B^{(2)}$. Because of the anomaly, the defect $D_3(M_3,B^{(2)})$ is not invariant with respect to background gauge transformations of $B^{(2)}$: only the following combination
\be
D_3(M_3,B^{(2)}) e^{i{2 \pi \over 2 N} p \int_{M_4} \mathcal{P}(B^{(2)})}, \qquad \partial M_4 = M_3,
\ee
is. As emphasized by Kaidi, Ohmori and Zheng \cite{Kaidi:2021xfk}, since this defect only depends on $M_4$ via the background, it is still a genuine defect of the theory. We are interested in gauging $\mathbb Z^{(1)}_N$, which leads to a new theory $\widetilde{\mathcal T}$. Upon such gauging we are promoting $B^{(2)}$ to a dynamical gauge field $b^{(2)}$. The resulting defect is no longer a genuine operator of $\widetilde{\mathcal T}$. To obtain a well-defined genuine topological defect of $\widetilde{\mathcal T}$, one needs to cancel the dependence of $D_3(M_3,b^{(2)})$ on $M_4$. When $\gcd(N,p)= 1$, this can be done with a straightforward generalization of the argument in \cite{Kaidi:2021xfk}: one can simply absorb the anomaly by stacking a copy of the minimal 3d TFT $\mathcal A_{N,-p}$ of \cite{Hsin:2018vcg} along $D_3(M_3,b^{(2)}) $ to obtain a new genuine defect in the $\widetilde{\mathcal{T}}$ theory:
\be
\mathcal N(M_3) = D_3(M_3,b^{(2)}) \otimes \mathcal A_{N,-p}(M_3,b^{(2)})\,.
\ee
Then the theory $\widetilde{\mathcal{T}}$ has a non-invertible $M$-ality defect $\mathcal N(M_3)$, with fusion rules determined from the properties of the 3d TFT $\mathcal A_{N,-p}$ and by the condensate of $\mathbb Z^{(1)}_N$ on $M_3$.\footnote{\ We refer our readers that are not familiar with the notion of condensate to the beautiful papers \cite{Gaiotto:2019xmp,Choi:2022zal,Roumpedakis:2022aik}.}

\medskip

If instead $\gcd(N,p)=k$, one can show that in this case the anomaly only involves the $\mathbb Z_{N/k}^{(1)}$ subgroup of $\mathbb Z_{N}^{(1)}$, assuming that $X_4$ is spin (and hence $\int_X \mathcal P(B^{2})$ is divisible by two).\footnote{\ Since we are working with supersymmetric theories in this paper, this is always the case.} Let us proceed demonstrating this explicitly. Since $\gcd(N,p)=k$, we have an exact sequence
\be
0 \to \mathbb Z_k \to \mathbb Z_N \to \mathbb Z_{N/k} \to 0,
\ee
and we can decompose
\be
B^{(2)} = {N \over k} B^{(2)}_k + B^{(2)}_{N/k},
\ee
where $B^{(2)}_k$ is a background for $\mathbb Z_k^{(1)}$ and $B^{(2)}_{N/k}$ for $\mathbb Z^{(1)}_{N/k}$. By massaging the Pontrjagin square, one can show that:\footnote{\ The relevant identities can be found in appendix \ref{app:potra}.}\be\label{eq:Nk}
\begin{aligned}
 {2 \pi p \over 2 N} \, \int_X \mathcal P(B^{(2)}) &=  {2 \pi p \over 2 N} \,  {N \over k} \int_X \mathcal P( B^{(2)}_k) + {2 \pi p \over 2 N} \int_X \mathcal P(B^{(2)}_{N/k}) +2 \pi \times (\text{integers})\,.\\
 \end{aligned}
 \ee
 The first term simplifies to 
 \be
 {2 \pi p \over 2 N} \,  {N \over k} \int_X \mathcal P( B^{(2)}_k) = \pi \ell \int_X \mathcal P( B^{(2)}_k), \qquad \text{where } p = k \ell\,.
 \ee
Since $\int_X \mathcal P( B^{(2)}_k) $ is even on spin manifolds, this term does not contribute to the anomaly. Thus,
\be
 {2 \pi p \over 2 N} \, \int_X \mathcal P(B^{(2)}) ={2 \pi p \over 2 N} \int_X \mathcal P(B^{(2)}_{N/k}) + 2 \pi \times (\text{integers}),\
\ee
and since
\be
{2 \pi p \over 2 N} = {2 \pi \ell \over 2 N/k},
\ee
we obtain that a non-trivial $B^{(2)}_{N/k}$ flux causes the anomaly
\be
{2 \pi p/k \over 2 N/k} \int_X \mathcal P(B^{(2)}_{N/k})\,.
\ee
Gauging the $\mathbb Z^{(1)}_k$ anomaly-free subgroup of the one-form symmetry, one obtains a theory with a symmetry group $\mathbb Z_{N/k}^{(1)}$ and the anomaly above. Gauging such  $\mathbb Z_{N/k}^{(1)}$, we obtain yet another new theory $\widetilde{\mathcal T}$, which has an $M$-ality defect obtained by
\be
\mathcal N_3(M_3) = D_3(M_3,b^{(2)}_{N/k}) \otimes \mathcal A_{N/k, -p/k}(M_3, b^{(2)}_{N/k})\,.
\ee
Notice that this defect is also genuine: the anomaly now can be absorbed by stacking a 3d TFT $\mathcal A_{N/k, -p/k}$ since $\gcd(N/k,p/k) = 1$. This implies that also in this case we obtain an $M$-ality defect, whose fusion is proportional to the $\mathbb Z^{(1)}_{N/k}$ condensate along $M_3$.

\section{Global structures from 6d -- a quick review}\label{sec:global6d}

\subsection{The 6d partition vector}
\begin{figure}
\begin{center}
\includegraphics[scale=0.3]{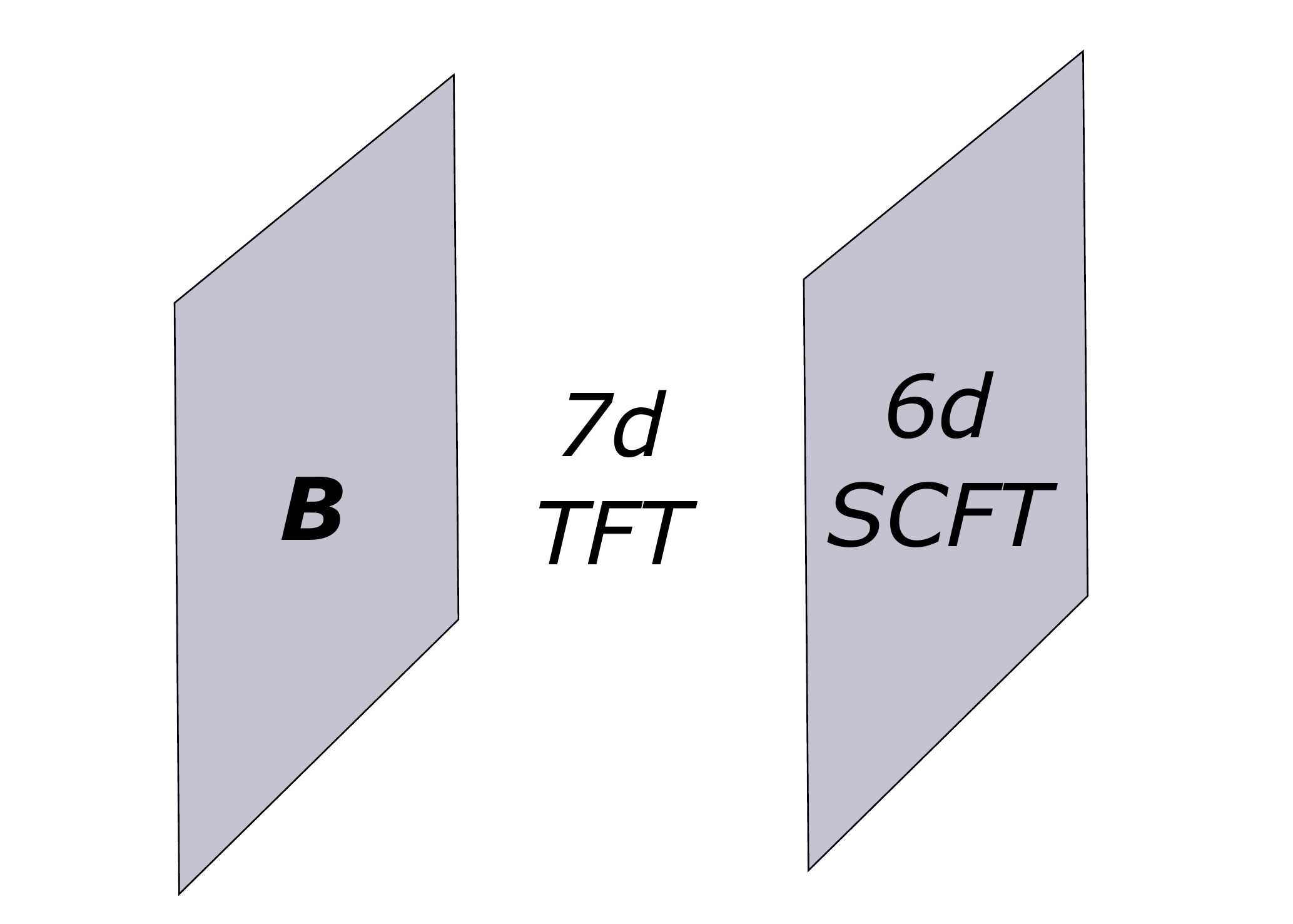}
\caption{Schematic picture of 6d SCFTs as relative field theories to a 7d TFT}\label{fig:7dtft}
\end{center}
\end{figure}

In this section we establish our notations and conventions by quickly summarising some of the features of 6d theories that will be useful below. We closely follow the presentation in \cite{Tachikawa:2013hya}. For the sake of brevity and clarity, in this paper we focus on the 6d (2,0) theories of type $A_{n-1}$. It is well-known that the $A_{n-1}$ 6d (2,0) theories are relative field theories \cite{Freed:2012bs, Witten:2009at}: given a compact closed torsionless six-manifold $Y$, the 6d (2,0) $A_{n-1}$ theory does not assign to it a complex number, a partition function, but rather a collection of partition functions, organized in a partition vector $| \mathcal Z(Y) \rangle$, which is an element of a Hilbert space. It is believed that the Hilbert space in question can be obtained from a non-invertible 7d TFT -- see figure \ref{fig:7dtft}: the 6d SCFT are understood as non-topological boundary conditions for such a theory on a 7d spacetime with $Y$ as a boundary. The Hilbert space can be characterised as a representation of a Heisenberg algebra of non-commuting discrete 3-form fluxes valued in $\mathbb Z_n$, the defect group of the 6d theory \cite{DelZotto:2015isa}. More precisely, the Heisenberg algebra in question is
\be \label{eq:quantumtorus}
  \Phi(a)\Phi(b) = e^{i\ev{a,b}} \Phi(b)\Phi(a), \qquad a,b\in H^3(Y,\mathbb Z_n),
\ee
where 
\be
\ev{a,b} = \frac{2\pi}{n} \int_Y a \cup b\,.
\ee
The Heisenberg algebra is also equipped with a canonical normal ordering prescription, which gives a group homomorphism from $H^3(Y,\mathbb Z_n)$ to the quantum torus in \eqref{eq:quantumtorus} \footnote{\ Indeed: $\Phi(b+a) = \Phi(b)\Phi(a)e^{i\ev{b,a}/2} =\Phi(a)\Phi(b)e^{i\ev{a,b}/2} = \Phi(a+b)$.}
\be\label{eq:norm}
\Phi(a+b) = \Phi(a)\Phi(b)e^{i\ev{a,b}/2} 
\ee
By the Stone-Neumann-Mackay theorem, for each choice of a maximally isotropic sublattice $\mathcal L$ of $H^3(Y,\mathbb Z_n)$ there is a unique ray in the Hilbert space $\mathcal H(Y)$, such that
\be
\Phi(\ell) |\mathcal L,0 \rangle = |\mathcal L ,0 \rangle \qquad\quad \forall \ell \in \mathcal L.
\ee
The rest of the basis elements of $\mathcal H(Y)$ are obtained from the elements $v \in \mathcal L^\perp \equiv H^3(Y,\mathbb Z_n)/\mathcal L$:\footnote{\ Here we are implicitly choosing a representative of $v\in \mathcal L^\perp$ inside $H^3(Y)$ -- of course the state  $|\mathcal L , v \rangle $ depends on this choice only up to a phase which can always be absorbed in a local counterterm.}
\be
|\mathcal L , v \rangle = \Phi(v)  |\mathcal L ,0 \rangle \qquad \forall \, v \in \mathcal L^\perp\,.
\ee
A crucial remark for us is that these are eigenvectors for the $\Phi(\ell)$ with $\ell \in \mathcal L$, with eigenvalues
\be
\Phi(\ell) |\mathcal L ,v \rangle = e^{i \ev{\ell,v}} |\mathcal L ,v \rangle\,.
\ee
Another consequence of the Stone-Neumann-Mackay theorem is that, given two maximal isotropic subgroups of $H^3(Y,\mathbb Z_n)$, say $\mathcal L$ and $\mathcal L^\prime$, the representations constructed in this way are isomorphic, meaning that there is an invertible linear transformation such that
\be
| \mathcal L^\prime , v^\prime \rangle = \sum_{v \in \mathcal L^\perp} {R_{\,v^\prime}}^v  |\mathcal L ,v \rangle \qquad \forall \, v ' \in {\mathcal L^\prime}^\perp.
\ee
For fixed $\mathcal L$, we can then write
\be
| \mathcal Z(Y) \rangle = \sum_{v \in \mathcal L^\perp} \mathcal Z_v(Y) |\mathcal L, v\rangle\,.
\ee
The coefficients $ \mathcal Z_v(Y)$ are the so-called 6d conformal blocks \cite{Witten:2009at}. Clearly, choosing a different maximally isotropic lattice $\mathcal L'$, one has another set of 6d conformal blocks  $\mathcal Z_{v'}(Y)$, but the two must be related. Indeed,
\be
| \mathcal Z(Y) \rangle = \sum_{v' \in {\mathcal L^\prime}^\perp}  \mathcal Z_{v'}(Y) |\mathcal L^\prime, v' \rangle =  \sum_{v' \in {\mathcal L^\prime}^\perp} \mathcal Z_{v'}(Y) \sum_{v \in \mathcal L^\perp} {R_{\,v^\prime}}^v  |\mathcal L ,v \rangle,
\ee
which implies that
\be
\mathcal Z_v(Y) =  \sum_{v' \in {\mathcal L^\prime}^\perp} \mathcal Z_{v'}(Y) {R_{\,v^\prime}}^v\,.
\ee
In order to extract values out of the partition vector, one can consider placing the 7d TFT on a finite interval times $Y$. On one side of the interval we have the relative 6d SCFT, on the other we insert a topological boundary condition, which in the figure we schematically denote \textbf{\textit{B}}. For example, we could be setting Dirichlet boundary conditions for all the 3-form fields, corresponding to the 3-form fluxes associated to the lattice $\mathcal L$. In this way we obtain a vector dual to $| \mathcal L, 0\rangle$. Hence,
\be\label{eq:cane}
Z_{\mathcal L} (Y) = \langle \mathcal L, 0 | \mathcal Z(Y) \rangle = \mathcal Z_0(Y).
\ee
Choosing different boundary conditions, corresponding to a different lattice, one obtains
\be
Z_{\mathcal L^\prime} (Y) = \mathcal Z_{0^\prime}(Y) = \sum_{v \in \mathcal L^\perp} \mathcal Z_v(Y) {(R^{-1})_{v}}^{0'} 
\ee
instead. This is the mechanism, which gives the 6d origin of the different partition functions, associated to possible global forms in lower dimensional field theories.

\subsection{6d origin of global structure of class $\mathcal{S}$ theories} \label{sec:glob-struct-class}

Class $\mathcal{S}$ theories are obtained by compactifying the 6d (2,0) theories on Riemann surfaces. In this work, to keep technicalities at a minimum, we focus on theories arising from Riemann surfaces $\Sigma_{g}$ of genus $g$ without punctures. For $g=1$, i.e. when the Riemann surface is a torus, the corresponding class $\mathcal{S}$ theories one obtains from this construction are the various $\mathcal{N}=4$ SYM with gauge algebra $\mathfrak{su}(n)$. For $g>1$, one obtains models with an S-duality frame, where they can be interpreted as conformal gaugings of $2g-2$ trinion $T_n$ theories \cite{Gaiotto:2009we} (see also \cite{Tachikawa:2015bga} for a thorough review), coupled to $3g -3$ gauge groups with $\mathfrak{su}(n)$ gauge algebras. 

\medskip

\noindent The various global structures are captured from 6d exploiting the conformal block expansion. Consider the 6d (2,0)
theory on a background $Y = \Sigma_{g} \times X$, where $X$ is the 4d spacetime. Assuming $X$ is compact, torsion-free, and that $H^1(X,\mathbb{Z})$ is trivial, from the Künneth formula and the universal coefficient theorem we obtain
\be
H^3(Y,\mathbb Z_n) \simeq H^1(\Sigma_g,\mathbb Z_n) \otimes H^2(X,\mathbb Z_n).
\ee
From our discussion above, for each fixed $n \geq 2$, the additional discrete data needed to fully specify the theory is a maximal isotropic lattice $L$ of $H^{1}(\Sigma_{g},\mathbb{Z}_n)$ with respect to the canonical pairing induced by the intersection pairing on $\Sigma_g$. Then
\be
\mathcal L = L \otimes H^2(X,\mathbb Z_n)
\ee
gives a maximal isotropic lattice of $H^3(Y,\mathbb Z_n)$ automatically. Different global forms for the 4d class $\mathcal S$ theories, corresponding to the surface $\Sigma_g$, are parametrized by such choices of maximally isotropic sublattices. Fixing one such sublattice, the fluxes in $\mathcal L^\perp$ parametrize the possible partition functions for the 4d theory with inequivalent 1-form symmetry backgrounds along $X$
\be
\begin{aligned}
Z_{\Sigma_g,L}(X,\xi) &= \langle \mathcal L , \xi | \,\mathcal Z(\Sigma_g \times X)\rangle \qquad \xi \in \mathcal L^\perp\,\\
&= \langle \mathcal L , 0 |\, \Phi(\xi)\, | \mathcal Z(\Sigma_g \times X)\rangle,
\end{aligned}
\ee
where we have used that $\langle \mathcal L , \xi | = \langle \mathcal L , 0 |\, \Phi(\xi)$ by definition.\footnote{\ Morally this should be a $\Phi^\dagger$, but since we always act on the left and never on the right, we won't pay attention to that.}

\section{Non-invertible symmetries from 6d}\label{sec:strata}

\subsection{0-form symmetry from mapping class group fixed points}
Four-dimensional $\mathcal N=2$ SCFTs have $\mathcal N=2$ conformal manifolds which can be described as the space of exactly marginal deformations modulo S-duality. For the theories of class $\mathcal{S}$ of interest for this paper the S-duality group has a beautiful geometrical description: the exactly marginal couplings of the $\mathcal N=2$ theory are identified with the complex structure coordinates of the Teichm{\"u}ller space of $\Sigma_g$, while the S-duality group is identified with the mapping class group. The conformal manifold then is given by $\mathcal M_g$, the moduli space of complex structures modulo the mapping class group.\footnote{\ This picture is slightly too naive because it does not keep track of the action of S-duality on the global structure of the theory. Including the global structure one obtains an extended conformal manifold. As we will see below, this remark has an important effect on our construction.}

Interestingly, the action of the mapping class group is not free: there can be points on the Teichm\"uller space that are fixed under the action of some subgroup of the mapping class group, and for that reason the conformal manifold is better characterized as a Deligne-Mumford stack. Whenever we have fixed points, naively one would claim that the theory corresponding to
such a point $\Sigma_{g}^*,$ which is stabilized by a subgroup $\mathbb{G}$ of the mapping class group, has an enhanced $\mathbb{G}^{(0)}$ symmetry. This statement is slightly too naive: it is sufficient to consider the simplest case, i.e. genus one, to understand the problem. The mapping class group in that case is $SL(2,\mathbb{Z})$ and it acts on the complexified coupling $\tau$, giving rise to the Montonen-Olive duality of $\mathcal{N}=4$ SYM. If that statement above were to be true, all $\mathcal N=4$ SYM theories would have an enhanced $\mathbb Z_2^{(0)}$ symmetry corresponding to the $S$ transformation
\be
\tau \to - 1 /\tau
\ee
at the self dual $\tau = i$. This statement is clearly false: $SU(n)$ is mapped to $PSU(n) = SU(n)/\mathbb Z_n$ by S-duality. This indicates there is a further requirement that needs to be imposed for $\mathbb{G}$ to give rise to an enhanced $\mathbb{G}^{(0)}$ symmetry: the resulting symmetry must respect the global structure of the theory. This requirement is transpart from the 6d perspective. An element $\phi$ of the mapping class group acts on the surface $\Sigma_{g}$ to give us a new surface $\phi(\Sigma_{g})$. It also induces an isomorphism between the cohomology of $\Sigma_{g}$ and that of $\phi(\Sigma_{g})$ by pull-back. As a result, it generically changes the lattice $\mathcal{L} \simeq L \otimes H^2(X,\mathbb Z_n)$ which defines the 1-form symmetry of the theory (and hence its global structure), as follows:
\be
\phi(\mathcal L) \equiv \phi_*(L) \otimes H^2(X,\mathbb Z_n)
\ee
In order for $\phi$ to truly generate a zero-form symmetry, it must fix not only $\Sigma_{g}$, but also the lattice $L$:
\be
\phi(\Sigma_g^*) = \Sigma_g^* \qquad \phi_*(L) = L \, \Rightarrow \, \phi(\mathcal L) = \mathcal L.
\ee
If these conditions are met, then
\be
Z_{\Sigma_g^*,L}(X,0) =  Z_{\phi(\Sigma_g^*),\phi(L)}(X,0)
\ee
and $\phi$ is indeed a symmetry of the theory.\footnote{\ When $\phi$ is not a symmetry it can give rise to more interesting effects, that give rise to intrinsic $K$-ality defects in the language of \cite{Kaidi:2022uux}. There one obtains a theory with a non-invertible symmetry starting from this setup as well, by coupling the 4d theory to an appropriate SPT to compensate to the given transformation \cite{Choi:2021kmx,Choi:2022zal}. We will describe these in a separate work.}

\medskip

For genus $g > 1$ the mapping class group has a rather complicated structure,\footnote{\ We refer to \cite{Farb2013APO} for an instructive review.} but in this paper we are only interested in subgroups of the mapping class group that stabilize some surface $\Sigma_{g}^*$: describing these is much simpler because the stabilizer in the mapping class group of a surface $\Sigma_{g}^*$ of genus $g > 1$ is isomorphic to the group of isometries of $\Sigma_{g}^*$.\footnote{\ For $g=1$ the isometry group always contains the group of translations and in addition there can be a discrete subgroup of isometries. This discrete subgroup is generically trivial, and is non-trivial only for the tori obtained by identifying the diagonally opposite sides of a square or a regular hexagon. In these cases it is isomorphic to the symmetries of the square or a regular hexagon, respectively.} For $g > 1$, the group of isometries is always finite, and like the finite symmetries of the torus, these isometries are a non-generic feature. The size of the group of isometries is bounded from above by $84(g-1)$, and the maximal order of a cyclic subgroup is $4g+2$.

Examples of surfaces with a $\mathbb{Z}_{4g+2}$ group of isometries can be constructed in a very analogous manner to the tori with discrete symmetry: we start with a regular $4g+2$-gon but, this time in hyperbolic space, and identify the diagonally opposite sides. This gives us a surface of genus $g$, which has rotations by an integer multiple of $\frac{2\pi}{4g+2}$ as its isometries (see e.g. figure \ref{fig:Decagon} as an example). Compactification on these surfaces can be exploited to obtain class $\mathcal{S}$ theories with a $\mathbb{Z}_{4g+2}^{(0)}$ enhanced symmetries.

\subsection{Reading off the mixed anomaly from 6d}

We now turn to the question of how we can use this technology as a diagnostic of mixed zero-form and one-form anomalies. In order to detect the 't Hooft anomaly, we act with our symmetry generator in presence of a non-trivial 1-form symmetry background. The simplest such background has a 6d avatar of the form $\beta \otimes v \in \mathcal L^\perp$ then
\be\label{eq:mfkr}
\begin{aligned}
Z_{\phi(\Sigma_g^*),\phi(L)}(X,\phi(\beta\otimes v)) &= Z_{\Sigma_g^*,L}(X,\phi(\beta\otimes v)) \\
& = \langle \mathcal L, 0 |\, \Phi(\phi(\beta\otimes v)) | \mathcal Z(\Sigma_g^* \times X)\rangle 
\end{aligned}
\ee
by construction. Now the key point is that $\phi(\beta \otimes v)$ is not necessarily an element of $\mathcal L^\perp$. For instance, consider the case
\be\label{eq:obvious}
\phi(\beta \otimes v) = \beta \otimes v + \alpha \otimes v \qquad \text{ where } \alpha \otimes v \in \mathcal L \ 
\ee
If that is the case, via the normal ordering prescription in equation \eqref{eq:norm}
\be
\Phi((\beta + \alpha)\otimes v)  \equiv \Phi(\beta \otimes v)\Phi(\alpha \otimes v) e^{i \tfrac{1}{2}\ev{\beta\otimes v,\alpha\otimes v}}
\ee
and \eqref{eq:mfkr} equals
\be
\begin{aligned}
&\langle \mathcal L, 0 |\, \Phi(\beta \otimes v)\Phi(\alpha \otimes v) e^{i \tfrac{1}{2}\ev{\beta\otimes v,\alpha\otimes v}} \, | \mathcal Z(\Sigma_g^* \times X)\rangle \\
&= \langle \mathcal L, 0 |\,\Phi(\alpha \otimes v) \Phi(\beta \otimes v) e^{i \tfrac{1}{2}\ev{\alpha\otimes v,\beta\otimes v}} \, | \mathcal Z(\Sigma_g^* \times X)\rangle \\
& = e^{i \tfrac{1}{2}\ev{\alpha\otimes v,\beta\otimes v}} \langle \mathcal L, 0 |\,\Phi(\beta \otimes v) | \mathcal Z(\Sigma_g^* \times X)\rangle \\
& =e^{i \tfrac{1}{2}\ev{\alpha\otimes v,\beta\otimes v}} Z_{\Sigma_g,L}(X,\beta\otimes v)
\end{aligned}
\ee
We obtain a mixed anomaly provided
\be\label{eq:conditionA}
e^{i \tfrac{1}{2}\ev{\alpha\otimes v,\beta\otimes v}} \neq 1.
\ee
Then \textit{mutatis mutandis} the argument of \cite{Kaidi:2021xfk} we can obtain a theory with a non-invertible symmetry by gauging the one-form symmetry associated to the choice of background  $\beta \otimes v \in \mathcal L^\perp$. In the next section we illustrate this mechanism in practice.
 
\section{Examples of applications}\label{sec:ex}
The technology developed in \ref{sec:strata} suggests the following process for finding mixed zero-form one-form anomalies and consequently non-invertible
duality defects:
\begin{enumerate}
    \item\label{item:1} Construct a class $\mathcal{S}$ theory with a discrete global symmetry using a surface $\Sigma_{g}$ with discrete isometries. For $g=0$
          this corresponds to the point $\tau=i$ which is self-dual under $S$ duality. For $g > 1$ we obtain theories with $\mathbb{Z}_{4g+2}$ $0-$form symmetry using the
          surface obtained by identifying the diagonally opposite edges of a regular hyperbolic polygon.
    \item\label{item:2} Find a maximal isotropic sublattice $\mathcal{L}$ of $H_{1}(\Sigma_{g},\mathbb{Z}_{n})$, which is invariant under the action of the discrete isometries
          constructed in the first step.
    \item\label{item:3} Find a background which under the action of the isometry changes by an element of the self dual lattice $\mathcal{L}$ constructed in the second step. If \eqref{eq:conditionA} holds, we have found a mixed anomaly, and
          consequently a non-invertible duality defect in the theory obtained by gauging the anomalous one-form symmetry.
\end{enumerate}
Below we present some examples where this strategy can be successfully carried out.

\subsection{The $\mathbb Z_2^{(0)} \times \mathbb Z_2^{(1)}$ mixed anomaly in $SO(3)_{-}$ at $\tau = i$ from 6d}\label{sec:an-example:-mixed}
As a warm-up and consistency check we now reproduce the mixed anomaly in $SO(3)_{-}$ using our formalism. The relevant surface in this case is a torus with complex
structure $\tau$. The $\mathbb{Z}_{2}$ valued cohomology of the torus is generated by $A$ and $B$ cycles with $A\cdot B = 1$ and has four elements 
\be
0,A,B,A+B\,.
\ee
The global form $SO(3)_{-}$ is obtained from the lattice $\mathcal{L} \simeq (A+B)\otimes H^2(X,\mathbb Z_n)$. In
this case $S$ duality acts by $\phi: \tau \mapsto -\frac{1}{\tau}$ and $A \mapsto B$, while $B \mapsto -A = A$ since $A+A = 0$. Since $\mathcal{L}$ is invariant under the exchange of $A$ and $B$, we
obtain a zero-form symmetry at the self dual point $\tau=i$. For $SO(3)_{-}$, the $S$-duality then implies
\begin{align}
  \label{eq:9}
  Z_{SO(3)_{-}}(-\tfrac{1}{\tau},X) = \langle \phi(\mathcal L) , 0 | \mathcal Z \rangle = \langle \mathcal L , 0 | \mathcal Z \rangle = Z_{SO(3)_{-}}(\tau,X)
  \end{align}
Now to detect the mixed anomaly, we wish to turn on a non-trivial background through a two cycle $v$ in the spacetime, hence we consider
\be
\begin{aligned}
Z_{SO(3)_{-}}(-\tfrac{1}{\tau},X,A\otimes v) &= \langle \phi(\mathcal L) , \phi(A\otimes v) | \mathcal Z \rangle \\
& = \langle \mathcal L , B\otimes v | \mathcal Z \rangle \\
& = \langle \mathcal L , (A + (A+B))\otimes v | \mathcal Z \rangle\\
&= e^{i \tfrac{1}{2}\langle A\otimes v, (A+B)\otimes v \rangle}\langle \mathcal L , A \otimes v  | \mathcal Z \rangle\\
&= e^{i \tfrac{1}{2}\langle A\otimes v, (A+B)\otimes v \rangle} Z_{SO(3)_{-}}(\tau,X,A\otimes v)\\
&= e^{i \tfrac{1}{2}\langle A\otimes v, B\otimes v \rangle} Z_{SO(3)_{-}}(\tau,X,A\otimes v)\\
&= e^{i \tfrac{\pi}{2} \mathcal P(v)}Z_{SO(3)_{-}}(\tau,X,A\otimes v)
\end{aligned}
\ee
which gives exactly the mixed anomaly at $\tau = i$ needed for generating a non-invertible duality symmetry along the lines discussed in \cite{Kaidi:2021xfk}.

\subsection{More $\mathcal{N}=4$ SYM examples from class $\mathcal{S}$}\label{sec:Nequal4}
As a first example let us consider $\mathcal{N}=4$ theory with gauge algebra $\mathfrak{su}(4n^{2})$ at $\tau=i$. The surface in this case is a square torus and
$S$ duality acts as rotation by $\frac{\pi}{2}$. The cohomology $H^1(T^2, \mathbb Z_{4n^2})$ is generated by the usual $A$ and $B$ cycles with $A\cdot B=1$ and relations $A^{4n^2} = B^{4n^2} = 1$ and has a total of $(2n)^{4}$ elements. We
choose as the discrete data the lattice $\mathcal L = L \otimes H^2(X,\mathbb Z_n)$ where $L$ is the sublattice of $H^1(T^2, \mathbb Z_{4n^2})$ generated by $2nA$ and $2nB$. The lattice $L \otimes H^2(X,\mathbb Z_n)$ is isotropic, since the intersection number of any two of its elements involves a factor of $4n^{2}$, which is zero in $\mathbb{Z}_{4n^{2}}$. Since it contains $(2n)^{2}$ elements, it is also
maximal. This choice corresponds to the gauge group $SU(4n^{2}) / \mathbb{Z}_{2n}$. Moreover, this
gauging is done without any discrete theta angle. The one-form symmetry group of this theory is $\mathbb{Z}_{2n} \times \mathbb{Z}_{2n}$.

The lattice $\mathcal{L}$ is manifestly self-dual under the $S$-duality which sends $A \to B$ and $B \to - A$. So the remaining task is to find a
subgroup of the one form symmetry which has a mixed anomaly with $S$-duality. For this we consider a background given by $(nA + nB)\otimes v \in \mathcal L^\perp$. Since $2(nA+nB) \otimes v \in \mathcal{L}$, this background is for a $\mathbb{Z}_{2}$ subgroup of the one-form symmetry. Now, under $S$-duality 
\be
\phi((nA + nB)\otimes v) = (nB - nA)\otimes v = (nA + n B) \otimes v  - 2 n A \otimes v
\ee
and the phase in equation \eqref{eq:conditionA} is
\be
{1\over 2} \langle(nA + n B) \otimes v, - 2 n A \otimes v \rangle = {1\over 2} {2 \pi \over 4 n^2} (- 2 n^2) \mathcal{P}(v)=  - {\pi \over 2} \mathcal{P}(v)
\ee
which exhibits a mixed anomaly between the symmetry $\mathbb Z^{(0)}_2$ at $\tau = i$ and the $\mathbb Z_2^{(1)}$ subgroup of $\mathbb Z_{2n}^{(1)} \times \mathbb Z_{2n}^{(1)}$ corresponding to the element $(nA + nB)\otimes v \in \mathcal L^\perp$.

\medskip

This example illustrates how it is possible to have further examples of $\mathfrak{su}(n)$ $\mathcal N=4$ SYM theories which exhibit a $\mathbb Z_2^{(0)} \times \mathbb Z_m^{(1)}$ mixed anomaly for a subgroup $\mathbb Z_m^{(1)}$ of the corresponding 1-form symmetry. Gauging such a subgroup one obtains a theory with a non-invertible duality defect.

\subsection{Higher genus examples with non-invertible symmetries}\label{sec:infinitelymany}
In the discussion so far, we have only generated examples of non-invertible duality defects. In this section we use the class $\mathcal S$ construction for higher genus Riemann surfaces to exhibit infintely many example which give rise to $M$-ality non-invertible zero-form symmetries.

\subsubsection{Geometry of Riemann surfaces $\Sigma_g^*$ with $\mathbb Z_{4g+2}$ isometry}

\begin{figure}
    \centering
    \includegraphics[width=0.6\textwidth]{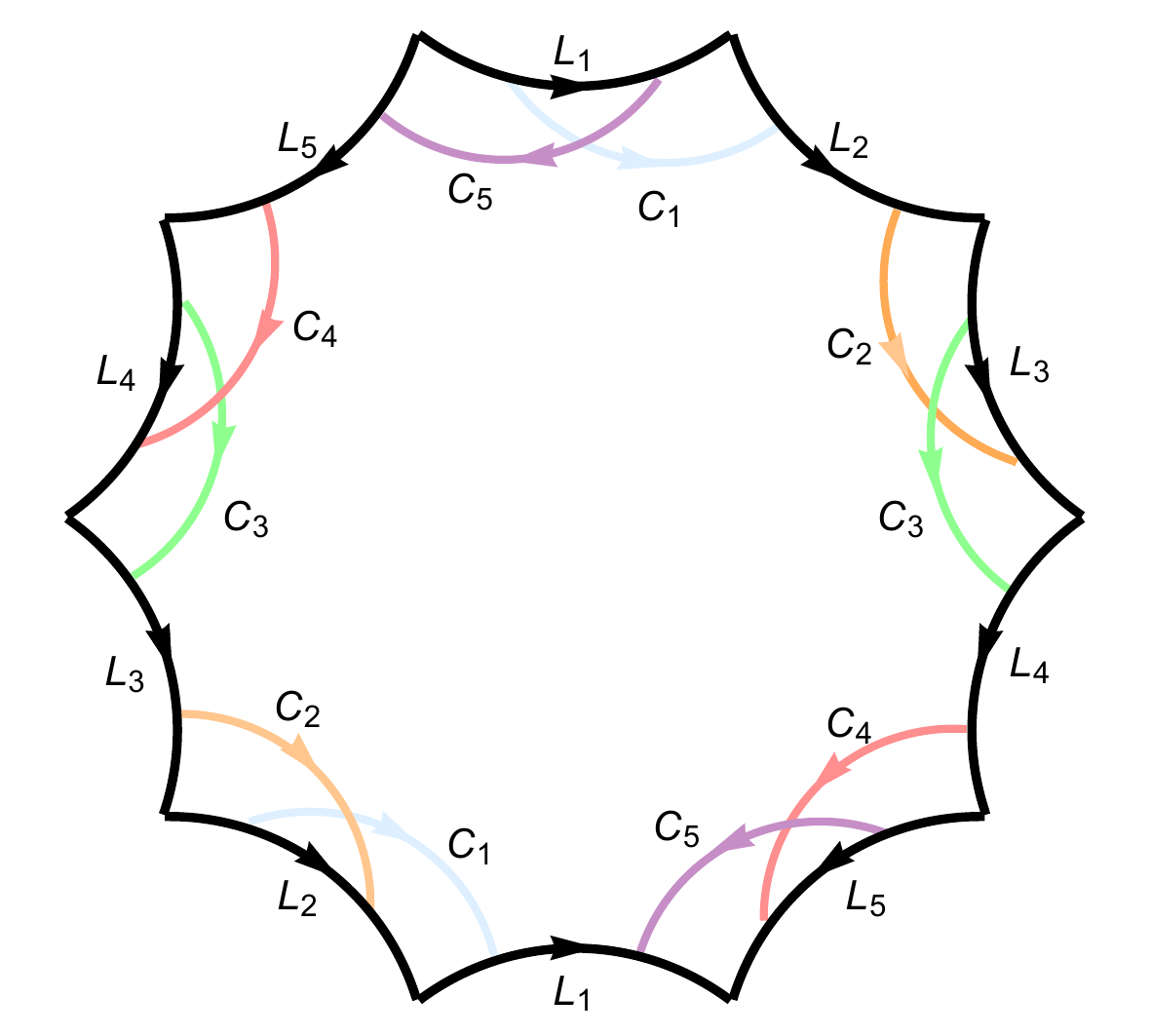}
    \caption{Decagon in the hyperbolic space, giving rise to a $g=2$ surface upon identification of the opposite sides. Coloured lines indicate homology cycles on the surface, among which any four are linearly independent.}\label{fig:Decagon}
\end{figure}

Let us begin by describing some general features of the genus $g$ surfaces $\Sigma_g^*$ which exhibit an isometry group $\mathbb{Z}_{4g+2}$. These can be obtained by identifying the opposite edges of a regular hyperbolic $4g+2$-gon. There is a (redundant) collection of 1-cycles $C_{1}, C_2, ..., C_{2g+1}$ which realise the $\mathbb{Z}_{4g+2}$ action as $C_i \mapsto C_{i+1}$ where the index is identified modulo $2g+1$ --- see figure \ref{fig:Decagon}.
The non-trivial intersection numbers are
\be
  C_{i} \cdot C_{i+1} = 1 \qquad 1 \le i \le 2g \qquad\text{and} \quad C_{2g+1} \cdot C_{1} = -1
\ee
and there is one relation between them:
\begin{align}
  \label{eq:extraCycle}
  C_{2g+1} = \sum_{i=1}^{2g} (-1)^{i}C_{i} ~.
\end{align}
A basis for $H_{1}(\Sigma_{g},\mathbb{Z}_{n})$ is given by $C_{i}$ with $1 \le i \le 2g$. The isometry $\phi$ rotating the polygon by $\frac{2\pi}{4g+2}$ acts on the independent cycles as,
\begin{align}
  \label{eq:21}
  \phi(C_{i}) &= C_{i+1} ~,  && 1 \le i \le 2g ~.
\end{align}
A generic class $\mathcal{S}$ theory obtained by compactifying 6d (2,0) theory with gauge algebra $\mathfrak{su}(n)$ on a genus $g$ surface $\Sigma_{g}$ has a
one form symmetry group of the same order as the order of a maximal isotropic sublattice of $H_{1}(\Sigma_{g},\mathbb{Z}_{n})$ i.e. $n^{g}$. We would like to find subgroups
of this one form symmetry, which have mixed anomaly with $\mathbb{Z}_{4g+2}^{(0)}$ zero-form symmetry generated by $\phi$.

\subsubsection{Infinitely many examples of $p^k$-ality defects}

Armed with the geometry described in the previous section we can now describe a generalisation of the examples found above whenever the genus satisfies
\be
2g+1 = p^{k}
\ee 
where $p$ is a prime. In this case we choose to work with 6d (2,0) $A_{n-1}$ theories that are such that
\be
n=p
\ee
so that we can view the cohomology group $H^1(\Sigma_g,\mathbb Z_p)$ as a $2g$-dimensional vector space over a finite
field $\mathbb F_p$. This has the advantage that we can use the Jordan decomposition to find subspaces invariant under $\phi$. 
\medskip

We proceed by computing the characteristic polynomial of $\phi$, i.e. $\det(\phi-x)$. We do this using the following trick: since for a prime $p$, 
\be
\binom{p^{k}}{m} = 0 \mod p
\ee
for $m \ne 0,\,p^k$, we obtain 
\be
(\phi-x)^{2g+1} = \phi^{2g+1} - x^{2g+1}\mod p.
\ee
Since $\phi^{2g+1} = -1$,
\begin{align}
  \label{eq:23}
  \det((\phi-x)^{2g+1}) = (\det(\phi-x))^{2g+1} = (-1)^{2g+1}(x^{2g+1}+1)^{2g+1} ~.
\end{align}
Next we use the fact that all the elements of the field $\mathbb F_{p}$ except $0$ form a multiplicative group of order $p-1$ so $x^{p^{k}} = x$. Hence,
\begin{align}
  \label{eq:24}
  \det(\phi-x) = (-1)^{2g+1}(x+1)^{2g+1} ~.
\end{align}
One can easily check that there is only one eigenvector of eigenvalue $-1$, by explicitly writing down the recurrence equation for the components of that
eigenvector, therefore  for each $\ell \le 2g$ there is a unique subspace invariant under the action of $\phi$ given by the kernel of $(\phi+1)^{\ell}$. Explicitly, there is a basis $D_{i}$ such that,
\begin{align}
  \label{eq:25}
  \phi(D_{1}) &= -D_{1} ~, \nonumber \\
  \phi(D_{i}) &= -D_{i} + D_{i-1} &&  1 < i \leq 2g~.
\end{align}
In terms of $C_{i}$, these new basis elements are given by,
\begin{align}
  \label{eq:26}
  D_{i} &= \sum_{j=0}^{2g-i}\binom{2g-i}{j}C_{j+1} ~.
\end{align}

The unique invariant subspace of dimension $d$ under the action of $\phi$ is given by the span of $D_{1} , \dots , D_{d}$. Specializing to $d=g$ we obtain the
unique invariant subspace that can determine a global structure. We just need to check if it is isotropic. To do that we note that $D_i=(\phi+1)^{2g-i}D_{2g}$. We then need to evaluate
\begin{equation}
  D_i \cdot D_j =  D_{2g}^T\left((\phi+1)^{2g-i}\right)^T \Omega (\phi+1)^{2g-j} D_{2g}.
\end{equation}
where we denoted $\Omega$ the pairing induced by the intersection form on $\Sigma_g$. The middle part of this expression can be reorganised as\footnote{\ Here we are using that $\phi^T \Omega \phi = \Omega$ since $\phi$ is an isometry, and therefore $\phi^T \Omega = \Omega \phi^{-1}$.}
\be
\begin{aligned}
\left((\phi+1)^{2g-i}\right)^T \Omega (\phi+1)^{2g-j}&=\Omega (\phi^{-1}+1)^{2g-i} (\phi+1)^{2g-j}\\
&=\Omega \phi^{-2g+i}(\phi+1)^{4g-i-j}.
\end{aligned}
\ee
Recalling that $(1+\phi)^{2g}D_{2g}=0$, we see that 
\be\label{eq:intersecati}
D_i \cdot D_j = 0 \qquad \forall \, i+j < 2 g+1.
\ee
Hence, the $g$-dimensional subspace spanned by $D_{1} , \dots  ,D_{g}$ is indeed isotropic, we denote it
\be
V = \text{span }(D_{1} , \dots  ,D_{g}) \subset H^1(\Sigma_g,\mathbb Z_p)\,.
\ee

\medskip

The theory $\mathcal T_{g,p}$ with global form corresponding to the maximal coisotropic lattice 
\be
\mathcal L_V = V \otimes H^2(X,\mathbb Z_n)
\ee
that we just identified has a $\mathbb Z_{4g+2}^{(0)}$ form enhancement of its symmetry at the self-dual point $\Sigma_g^*$ of $\mathcal M_g$ with $\mathbb Z_{4g+2}$ isometry.

\medskip

To exhibit an anomaly we can consider a background of the form $v \otimes D_{g+1}$, we have
\be
\phi(D_{g+1}) = - D_{g+1} + D_g \qquad\text{and}\qquad \phi^2(D_{g+1}) = D_{g+1} - 2 D_{g} + D_{g-1} 
\ee
while the action of $\phi$ does not have the form $\phi(\beta \otimes v) = \beta \otimes v + \alpha \otimes v$, the action of $\phi^2$ indeed does, and we have
\be
\phi^2(\beta \otimes v) = \beta \otimes v + \alpha \otimes v \qquad\text{with}\begin{cases} \beta = D_{g+1}\\ \alpha = - 2 D_g + D_{g-1}\end{cases}
\ee
therefore we see that the $\mathbb Z_{2g+1}^{(0)}$ subgroup of  $\mathbb Z_{4g+2}^{(1)}$ generated by $\phi^2$ can have a mixed anomaly with a subgroup of the $(\mathbb Z_n^g)^{(1)}$ one-form symmetry for all these models. It is easy to see that there is an anomaly, since the phase in equation \eqref{eq:conditionA} is\footnote{\ Here we are using that $D_{g+1} \cdot D_{g-1} = 0$ by \eqref{eq:intersecati} and we also must have that $D_{g+1} \cdot D_g$ must be non-zero: if it was zero, this would violate the maximality of the isotropic sublattice generated by $D_1,...,D_g$.}
\be
{1\over 2} \langle \alpha \otimes v, \beta\otimes v  \rangle = {\pi \over n}  (D_{g+1} \cdot D_g )\int_X \mathcal P(v) \,.
\ee
 Then we produce a mixed anomaly of the form discussed in section \ref{sec:MALITY}, between the $\mathbb Z_{2g+1}^{(0)}$ subgroup of the duality symmetry and the $\mathbb Z_p^{(1)}$ subgroup of the one-form symmetry $((\mathbb Z_{p})^g)^{(1)}$ which corresponds to the subspace of $\mathcal L^\perp$ associated to $D_{g+1} \otimes H^2(X,\mathbb Z_n)$. It is the condensate of this subgroup which features in the resulting fusion algebra.

\medskip

These anomalies are interesting in their own right are the smallest examples of mixed anomalies at a given genus when $2g+1 = p^{k}$ for some prime $p$. When $2g+1$ has more than one prime factors, we again need to look at the corresponding prime fields for the smallest examples. We expect these mixed anomalies, obtained when the homology
of the Riemann surface is a vector space over a finite field, to be building blocks for more general mixed anomalies. We plan to return to a more systematic study of these phenomena in a future work.

\section{Conclusions}\label{sec:conclusions}

In this note we have initiated the study of non-invertible defects in 4d $\mathcal{N}=2$ theories of class $\mathcal{S}$. We have exploited the observation that such defects can be constructed, whenever a mixed anomaly between a zero-form symmetry and a one-form symmetry is present. Starting from this point, and using the insights coming from the 6d perspective, we were able to recover some known cases of 4d $\mathcal{N}=4$ theories, possessing non-invertible defects, as well as to provide a bunch of new examples, coming from the higher-genus surfaces. In particular, we provide an infinite family of examples with increasing $M$-ality and $M=p^k$.

\medskip

Our results presented here are just the tip of the iceberg, and several very natural directions for further explorations can be identified. First, a more systematic exploration of the $M$-ality defects we predict is interesting. We are currently studying the resulting fusion algebras, and we will present them in a follow-up of this short note. Second, so far we have considered only theories of $A_n$ type. At the same time, it was observed \cite{Bhardwaj:2022yxj} that for certain choices of the global structure also gauge theories with the gauge algebra of $D_n$ type may have non-invertible defects in the operator spectrum.\footnote{\ Fusion algebras arising from gauging an outer automorphisms acting non-trivially on the one-form symmetries \cite{Bhardwaj:2022yxj} can be easily realized geometrically. We present some examples with these features in appendix \ref{app:othermoregeneral}, as an appetizer.} This is a good motivation to extend our analysis to the class $\mathcal{S}$ theories of $D_n$ and $E_n$ types. Third, we were concentrating exclusively on the theories obtained from compactifications of 6d $(2,0)$ theories on surfaces without punctures. It would certainly be interesting to extend the scope of examples by considering surfaces with punctures, regular or irregular ones. This exercise is interesting because in these examples we expect to be able to exhibit generalizations of symmetry defects corresponding to non-abelian finite zero-form symmetry groups arising from fixed points of the mapping class group.

\medskip

It should also be mentioned that, while we have mostly been focused on the duality defects coming from the aforementioned mixed anomalies, there is another tool based on the self-duality a theory might obey and the corresponding Kramers-Wannier duality defects \cite{Choi:2021kmx}. Exploiting our techniques it is easy to exhibit a broad scope of examples where the first method does not apply, while the second one is quite fruitful.

\medskip

Finally, a potentially interesting class of examples to which our methods also apply is provided by compactifications of 6d $(1,0)$ theories down to 4d (see
e.g.
\cite{Ohmori:2015pua,DelZotto:2015rca,Cecotti:2015hca,Ohmori:2015pia,Ohmori:2015tka,Morrison:2016nrt,Razamat:2016dpl,Bah:2017gph,Kim:2017toz,Razamat:2018gro,Kim:2018bpg,Kim:2018lfo,Zafrir:2018hkr,Ohmori:2018ona,Pasquetti:2019hxf,Razamat:2019ukg,Baume:2021qho,Distler:2022yse,Heckman:2022suy,Giacomelli:2022drw}
for a (partial) list of references on the subject). In this context the role of the Heisenberg algebra we discuss in this paper is played by the corresponding
Heisenberg algebra arising from the 6d defect group of the 6d (1,0) SCFT \cite{DelZotto:2015isa,GarciaEtxebarria:2019caf}.

\section*{Acknowledgements}

We warmly thank Justin Kaidi for carefully reading our draft, an illuminating email exchange, and discussions during the global categorical symmetry conference held at PI in June 2022. We thank Lea Bottini, Luca Cassia, Luis Diogo, Iñaki García Etxebarria, Jonathan Heckman, Usman Nasser,  Jian Qiu, and Sakura Schäfer Nameki for discussions. The work of MDZ and AH has received funding from the European Research Council (ERC) under the European Union’s Horizon 2020 research and innovation programme (grant agreement No. 851931). MDZ also acknowledges support from the Simons Foundation Grant \#888984 (Simons Collaboration on Global Categorical Symmetries). This research was supported in part by Perimeter Institute for Theoretical Physics. Research at Perimeter Institute is supported by the Government of Canada through the Department of Innovation, Science and Economic Development and by the proviince of Ontario through the Ministry of Research and Innovation.

\appendix

\section{On the proof of equation \eqref{eq:Nk}}\label{app:potra}
In order to prove \eqref{eq:Nk} one needs to start from the following property of the Pontryagin square operation, which follows from the fact that it is a quadratic form
\be
\int_X \mathcal P(A+B) =  \int_X \mathcal P(A) + \int_X \mathcal P(B) + 2  \int_X A \cup B\,\,,
\ee
and applying it recursively one can show that
\be
\int_X \mathcal P(\ell A) = \ell \int_X \mathcal P(A) + (2 \times \text{integer})  \int_X A \cup A
\ee
Indeed
\be
\begin{aligned}
\int_X \mathcal P(\ell A) &= \int_X \mathcal P(A) + \int_X \mathcal P((\ell-1)A) + 2 (\ell-1)  \int_X A \cup A\\
&= 2 \int_X \mathcal P(A) + \int_X \mathcal P((\ell-2)A) + 2 (\ell-2)  \int_X A \cup A + 2 (\ell-1)  \int_X A \cup A\\
& = \cdots \\
&= \ell \int_X \mathcal P(A) + (2 \times \text{integer})  \int_X A \cup A
\end{aligned}
\ee

\section{Categorical symmetries from outer automorphisms}\label{app:othermoregeneral}

Another fruitful strategy to produce non-invertible fusion rules, that goes beyond condensation and higher gauging \cite{Gaiotto:2019xmp,Choi:2022zal,Roumpedakis:2022aik}, arises when we are gauging an outer automorphism of the theory which is acting non-trivially on the one-form symmetry \cite{Bhardwaj:2022yxj}.

\medskip

The class of models we are considering in this paper exhibits a natural outer automorphism: consider a degeneration of $\Sigma_g$ such that it becomes a sphere with $g$ handles located symmetrically. This arrangement gives rise to an action of the group $\mathbb Z_g$ by cyclically permuting the various handles of $\Sigma_g$ with one another. This is not a duality defect, rather an outer automorphism of the SCFT, but we can consider gauging such subgroup. Since the latter acts on the various factors of the 1-form symmetry of the system associated to the various handles, upon gauging we obtain non-invertible 2-form symmetries that have a twisted-sector like fusion.

\medskip

Labeling $D_{i}$ the $i$-th codimension two topological surface defect corresponding to the $i$-th factor of the one-form symmetry $((\mathbb Z_n)^g)^{(1)}$, the gauged theory will have non-invertible codimension 2 defects of the form
\be
\mathcal N_{\oplus_j (i_1^{k_1},...,i_g^{k_g})_j} = \big[\bigoplus_j D_{i_{1,j}}^{k_{1,j}} \otimes D_{i_{2,j}}^{k_{2,j}} \otimes \cdots \otimes D_{i_{g,j}}^{k_{g,j}}\big]
\ee
which are labeled by gauge invariant orbits with respect to the $\mathbb Z_g^{(0)}$ action.

\medskip

Consider for simplicity the case $g=2$, then we have
\be
\mathcal N (M_2) = [(D_{1}(M_2)\otimes 1) \oplus (1 \otimes D_{2}(M_2))]
\ee
with a fusion algebra of the form
\be
\mathcal N(M_2) \times \mathcal N (M_2) = \mathcal N_{(1^2,0)\oplus (0,1^2)} \oplus \mathcal N_{(1,2)}
\ee
where
\be
\mathcal N_{(1^2,0)\oplus (0,1^2)}  = [(D_{1}^2(M_2) \otimes 1) \oplus (1 \otimes D_{2}^2(M_2) )]
\ee
and 
\be
\mathcal N_{(1,2)} = [D_{1}(M_2) \otimes D_{2}(M_2)]\,.
\ee
The study of these structures can be carried out along the lines discussed in \cite{Bhardwaj:2022yxj}, and we plan to attack them using the methods discussed above in a subsequent work. We report of their existence in this appendix because these give a slightly different example of a 6d origin for 4d non-invertible symmetry defects thus complementing the results presented in this first exposition.

\bibliographystyle{ytphys}
\bibliography{dualitydefects2.bib}

\end{document}